\documentclass[aps,pra,twocolumn]{revtex4}
\usepackage{graphicx}
\usepackage{bm}
\usepackage{amsmath,amssymb,amsthm,dsfont,bm}
\usepackage{color}
\usepackage{wasysym}
\usepackage{physics}
\usepackage{ulem}
\usepackage{appendix}
\usepackage[dvipsnames]{xcolor}

\usepackage[english]{babel} 
\usepackage[utf8]{inputenc} 
\usepackage{hyperref}
\usepackage{cancel}

\begin{document}
\preprint{}
\title{Kirkwood-Dirac distributions in classical optics}
\author{Alfredo Luis}
\affiliation{Departamento de \'Optica, Facultad de Ciencias F\'{\i}sicas, Universidad Complutense, 28040 Madrid, Spain}

\author{Lorena Ballesteros Ferraz}
\affiliation{Laboratoire de Physique Théorique et Modélisation (LPTM) - UMR CNRS 8089, CY Cergy Paris Université, Cergy-Pontoise, France}

\date{\today}

\begin{abstract}
We develop a comprehensive analysis of the Kirkwood-Dirac distributions in classical optics, revealing their deep connection with optical coherence as fundamental concept in optics. From their very definition, the Kirkwood-Dirac distributions emerge as generalized mutual coherence functions involving two different bases instead of just one. This perspective provides a unified interpretation of the so-called anomalous values, that are complex and negative values, as direct manifestations of coherence. We show that this interpretation consistently applies across all field variables considered in this work, including polarization, interference and wave propagation. Furthermore, we propose diverse methods of experimental determination of these distributions based on interference, in full agreement with their coherence-based interpretation. \\
\end{abstract}
\maketitle

\section{Introduction}

The statistical description of physical systems differs profoundly between classical and quantum theories. In classical physics, joint probability distributions provide a complete characterization of correlations between observables. In contrast, in quantum mechanics the noncommutativity of observables prevents the construction of genuine joint distributions for incompatible quantities~\cite{Dirac45}. This fundamental limitation has motivated the introduction of quasiprobability distributions~\cite{Wigner32,Kirkwood33,CahillGlauber69}, which extend classical statistical concepts while allowing for features such as negativity or complex values.

Among these constructions, here we will focus on the one originally introduced in quantum optics in the seminal works of Kirkwood~\cite{Kirkwood33} and Dirac~\cite{Dirac45}. The Kirkwood–Dirac distribution provides a formal joint representation of two observables associated with different bases. For a quantum state described by a density operator $\rho$, the Kirkwood–Dirac distribution is defined as $K(a,b)=\langle a|\rho|b\rangle \langle b|a\rangle$, and yields the correct marginals while, in general, taking complex or negative values. Unlike other quasiprobability distributions, such as the Wigner function, the Kirkwood–Dirac distribution is intrinsically complex and directly encodes the interplay between state coherence and the incompatibility of observables.

The Kirkwood–Dirac distribution has attracted renewed interest in recent years, particularly in connection with weak measurement theory~\cite{AAV88}. Within this framework, it provides an operational and experimentally accessible description of quantum systems~\cite{Wagner23}, and is closely related to weak values, which can be understood as conditional averages of the Kirkwood–Dirac quasiprobability distribution. This connection reveals that the Kirkwood–Dirac distribution underlies a wide range of phenomena including quantum interference, contextuality, and measurement back-action~\cite{Lostaglio23, Wagner23, ArvidssonShukur24}. Its complex and negative values are commonly interpreted as signatures of nonclassical behavior, especially when associated with incompatible observables and anomalous weak values~\cite{Ferrie11, HeFu23}.

Experimentally, the Kirkwood–Dirac distribution has been reconstructed in quantum systems using interferometric and auxiliary‑system schemes that provide access to both the real and imaginary components of the distribution~\cite{Hernandez-Gomez24,Wagner23}. These approaches go beyond conventional tomography to directly probe complex quasiprobability statistics associated with noncommuting observables. In the context of classical optics, early implementations based on heterodyne detection have demonstrated that the Kirkwood–Dirac distribution can also be accessed experimentally~\cite{LRBWT99,BSL10}, revealing a formal parallel between quantum states and classical coherence functions.

This parallel is rooted in the well-established analogy between quantum density operators and classical mutual coherence functions. In classical optics, the statistical properties of partially coherent fields are described by correlation matrices that share the same mathematical structure as quantum density matrices. This correspondence suggests that the Kirkwood–Dirac distribution may be naturally interpreted, in both domains, as a generalized coherence function involving two distinct modal bases.

In this work, we exploit this viewpoint to develop a unified treatment of the Kirkwood–Dirac distribution in classical optics, with particular emphasis on its operational accessibility and physical interpretation (Section~\ref{section_2}). We present two experimentally feasible schemes that allow for the observation of the Kirkwood–Dirac distribution in classical optical systems. These schemes are based on simple interferometric configurations and do not rely on heterodyne detection (Sections~\ref{section_3}~and~\ref{section_5}). Specifically, we address the polarization and space-angular degrees of freedom, showing how the Kirkwood-Dirac distribution can be reconstructed from measurable interference signals. 

A central question concerns the interpretation of the anomalous features of the Kirkwood–Dirac distribution, namely its complex values and the possible negativity of its real part. While these features are often regarded as hallmarks of nonclassicality in quantum systems, their appearance in classical optical fields calls for a more nuanced interpretation. In this work, we show that such anomalous values arise not only from coherence, off-diagonal correlations, but also from the interplay between population imbalance and the geometric relation between the chosen bases~(Sections~\ref{section_3}~and~\ref{section_5}). This leads to a reinterpretation of these features as signatures of generalized coherence rather than exclusively quantum behaviour.

Finally, we establish connections between the Kirkwood–Dirac distribution and experimentally accessible joint distributions arising from noisy simultaneous measurements (Section~\ref{section_4}), as well as its relation to the Wigner function in both continuous and discrete variable settings (Section~\ref{section_6}) and finally we conclude in Section~\ref{section_8}. These results reinforce the view of the Kirkwood–Dirac distribution as a refined descriptor of coherence and interference, providing a unified framework that bridges classical and quantum optics.

\section{Quantum and classical Kirkwood-Dirac distributions}\label{section_2}

The definition of the Kirkwood-Dirac distribution in the quantum case is
\begin{equation}
 K (a,b)= \langle a |\rho| b \rangle\langle b |a \rangle ,
\end{equation}
where $\rho$ is the density matrix, and $|a\rangle$, $|b \rangle$ are two different complete orthonormal bases with 
\begin{equation}
    \sum_a |a\rangle \langle a | = \sum_b |b \rangle \langle b | =\mathcal{I} ,
\end{equation}
where $\mathcal{I} $ is the identity. The distribution is normalized and has correct marginals
\begin{equation}
    \sum_b K(a,b) = \langle a |\rho| a\rangle, \quad \sum_a K(a,b) = \langle b |\rho| b\rangle, 
\end{equation}
and then
\begin{equation}
    \sum_{a,b} K(a,b) =1 .
\end{equation}
In passing, we can notice another interesting property. Expressing 
\begin{equation}
\label{hK}
    K(a,b) = {\rm tr} \left [ \rho \hat{K} (a,b) \right ], \qquad \hat{K} (a,b)= |b \rangle \langle b | a \rangle \langle a | ,
\end{equation}
we have
\begin{equation}
    {\rm tr} \left [ \hat{K} (a,b)\hat{K}^\dagger (a^\prime,b^\prime)\right  ] = \left | \langle a | b \rangle \right |^2 \delta_{a,a^\prime} \delta_{b,b^\prime}
\end{equation}
where $\delta_{j,k}$ is the Kronecker delta.

In general $K(a,b)$ can be complex and take negative values. To some extent $K(a,b)$ is one of the best options we have in quantum mechanics to approach the joint distribution of incompatible observables. Then its negative  and complex values can be taken as signatures of non classical behavior. 

Beyond its quantum value as nonclassical witness, in this work we intend to address its relation with coherence, both in the quantum and classical domains. In recent times, the deep relation between quantumness and coherence is being extensively developed~\cite{AL22}. In this regard, we can appreciate that the core of $K(a,b)$ are nondiagonal matrix elements $\langle a |\rho| b \rangle$, actually involving two different bases, so this is a kind of generalization of the idea of coherence beyond a single computational basis. 

Let us move to classical optics. The classical analogs of the density matrices are mutual coherence functions and polarization matrices, we will denote as $\Gamma$. This points to  a deep coherence meaning to the Kirkwood-Dirac distribution. Let us briefly recall this fruitful connection between density matrices and mutual coherence functions. 

For definiteness we will proceed in the space-frequency domain, and paraxial approximation, so that time essentially represents length along the axis and coordinates $\mathbf{r}=(x,y)$ describe transversal planes. Let us express a given field state in a proper modal expansion 
\begin{equation}
\label{mea}
    \mathbf{E} (\mathbf{r},t) = \sum_a E_a \bm{\mathcal{U}}_a (\mathbf{r},t) ,
\end{equation}
where $\bm{\mathcal{U}}_a (\mathbf{r},t)$ are the elements of a complete set of orthogonal modes for the problem at hand as the corresponding deterministic solutions of the Maxwell equations with the desired properties, and $E_a$ are the corresponding complex amplitudes as random field variables. Similarly we may have equally well considered another set of modes  $\bm{\mathcal{U}}_b (\mathbf{r},t)$
\begin{equation}
\label{meb}
    \mathbf{E} (\mathbf{r},t) = \sum_b E_b\bm{\mathcal{U}}_b (\mathbf{r},t) ,
\end{equation}
and in both cases
\begin{equation}
    E_{a,b} = \int d^2 \mathbf{r} \; \bm{\mathcal{U}}^\ast_{a,b} (\mathbf{r},t)  \mathbf{E} (\mathbf{r},t) .
\end{equation}
The parallel with quantum Hilbert space is rather enlightening, and we can express such field state in Eqs.~(\ref{mea})~and~(\ref{meb}) as 
\begin{equation}
    | E \rangle  = \sum_a E_a |a \rangle = \sum_b E_b |b \rangle .
\end{equation}
The random nature of the $E_{a,b}$ amplitudes may be taken formally into account by some probability distribution $P(E)$ for the vector $|E \rangle$. On these conditions, the classical analogs of the quantum matrix elements are given by the following ensemble averages
\begin{equation}
     \langle a |\rho| b \rangle = \Gamma (a,b) = \left \langle E_a E^\ast_b \right \rangle = \int d E P(E)  E_a E^\ast_b .
\end{equation}
In the Hilbert-space language this is to say that the classical density matrix is 
\begin{equation}
\label{enav}
\rho \rightarrow \Gamma = \int d E P(E) |E \rangle \langle E | ,
\end{equation}
so that $\Gamma$ is Hermitian $\Gamma^\dagger = \Gamma$ and positive semidefinite, $\Gamma \ge 0$. Then we have
\begin{equation}
    \Gamma (a,b) = \langle a | \Gamma | b \rangle ,
\end{equation}
and finally
\begin{equation}
    K (a,b) =  \Gamma (a,b) \langle b | a \rangle ,
\end{equation}
where
\begin{equation}
    \langle b |a \rangle = \int d^2 \mathbf{r} \; \bm{\mathcal{U}}_b^\ast (\mathbf{r},t) \bm{\mathcal{U}}_a (\mathbf{r},t) .
\end{equation}
These expressions show the deep coherence meaning of Kirkwood-Dirac distributions in the classical and quantum cases.

We must stress that in the classical domain the Kirkwood-Dirac distribution lacks any statistical connection other than being ensemble averages. But it may still be regarded as providing information about how much light is there having properties $|a\rangle$ and $|b\rangle$ simultaneously. There are cases in which the meaning of this is clear, say how much light at point $a$ propagates in direction $b$, as we shall see regarding the space-angular distribution. But in other cases the subject is rather cumbersome, as it is the case of polarization, where is difficult to imagine what means to have light in the polarization states $|a\rangle$ and $|b\rangle$ at the same time. In this regard, the quantum case is of no much help for that matter. On the other hand, its coherence connection is always there, and is always clear, both in the quantum and classical case, as we shall see in what follows.

\section{Polarization distribution}\label{section_3}
In this section, we introduce a simple interferometric scheme for measuring the Kirkwood-Dirac distribution associated with the polarization degree of freedom. In this case, modes represent just transversal polarization states, say $|a\rangle$ may represent vertical and horizontal linear polarization, while $| b \rangle$ may describe dextro and levo circular polarization, being these vectors the eigenstates of the corresponding Pauli matrices. The transverse polarization of a fully polarized classical beam can be represented by a complex two-dimensional vector $|E\rangle$. A general pure polarization state may be parameterized on the Poincaré sphere by two angles,
\begin{equation}
\label{E}
    |E\rangle = \begin{pmatrix}
        E_x \\
        E_y
    \end{pmatrix} =
    \begin{pmatrix}
        \cos\frac{\theta}{2} \\
        e^{i\varphi}\,\sin \frac{\theta}{2}
    \end{pmatrix},
\end{equation}
where $\theta$ and $\varphi$ determine the relative amplitudes and phase between the linear polarization components. 

When the statistical fluctuations of the field $|E\rangle$ are taken into account, the resulting light may no longer be fully polarized, and the polarization state is described by the $2\times 2$ complex coherence or polarization matrix $\Gamma$, expressed, for example, in the basis of the linear components $E_{x,y}$ along axes $X,Y$
\begin{equation}
\label{MG}
   \Gamma =  \begin{pmatrix} \Gamma_{1,1} & \Gamma_{1,2}   \\
       \Gamma_{2,1} & \Gamma_{2,2}
    \end{pmatrix} = \begin{pmatrix} \langle | E_x |^2  \rangle & \langle E_x E_y^\ast \rangle  \\
       \langle E^\ast_x E_y \rangle & \langle | E_y |^2 \rangle 
    \end{pmatrix} ,
\end{equation}
This is a suitable classical counterpart of the density matrix $\rho$. In order to make the analogy more complete we may consider units in which the total intensity is unity $\Gamma_{1,1}+\Gamma_{2,2} = 1$. 

Throughout this article, it will be rather useful to express $\Gamma$ in a suitable basis of matrices, such as the Pauli matrices $\hat\sigma_{x,y,z}$ and the $2 \times 2$ identity $\hat\sigma_0 $
\begin{equation}
\label{GSp}
    \Gamma = \frac{1}{2} \left ( \hat\sigma_0+ s_x \hat\sigma_x + s_y \hat\sigma_y + s_z \hat\sigma_z \right ) .
\end{equation}
where $\bm{s}$ are actually the Stokes parameters,
\begin{equation}
    s_j = \mathrm{tr} \left ( \Gamma \hat\sigma_j \right ) ,\qquad j=x,y,z ,
\end{equation}
that for the polarization state~(\ref{E}) becomes:
\begin{equation}
s_x= \sin \theta \cos \varphi, \quad s_y= \sin \theta \sin \varphi, \quad s_z= \cos \theta . 
\end{equation}

The deviation of $\Gamma$ from idempotency, that is, the extent to which $\Gamma^2 \neq \Gamma$, quantifies the degree of polarization of the beam, in direct analogy to the mixedness of quantum states. Fully polarized light corresponds to an idempotent (rank-one) matrix, whereas partially polarized light corresponds to a non-idempotent mixture of polarization components.

Let us proceed further to obtain the Kirkwood-Dirac distribution referred to two arbitrary bases of polarization, say
\begin{equation}
    |a \rangle \langle a | = \frac{1}{2} \left ( \hat\sigma_0+ a \bm{s}_A\cdot \hat{\bm{\sigma}} \right ) , \quad |b \rangle \langle b | = \frac{1}{2} \left ( \hat\sigma_0+ b\bm{s}_B\cdot \hat{\bm{\sigma}} \right ) ,
\end{equation}
where $a,b=\pm 1$ and $\bm{s}_A^2 = \bm{s}_B^2 = 1$ since they are fully polarized pure states. Using well-known properties of Pauli matrices we have that the $\hat{K} (a,b)$ in Eq.~(\ref{hK}) becomes:
\begin{eqnarray}
&   \hat{K} (a,b)= \frac{1}{4} \left [ \left ( 1+ ab \bm{s}_A \cdot \bm{s}_B  \right ) \hat\sigma_0 \right .& \nonumber \\ & & \nonumber \\
&\left .  + \left ( a \bm{s}_A + b \bm{s}_B +i ab  \bm{s}_A \times \bm{s}_B \right )\cdot \hat{\bm{\sigma}}  \right ] & 
\end{eqnarray}
and then 
\begin{eqnarray}
\label{Kab}
&   K(a,b)= \frac{1}{4} \left [ 1+ ab \bm{s}_A \cdot \bm{s}_B   \right .& \nonumber \\ & & \nonumber \\
&\left .  + \left ( a \bm{s}_A + b \bm{s}_B +i ab  \bm{s}_A \times \bm{s}_B \right )\cdot \bm{s} \right ] & .
\end{eqnarray}
\color{black}

\begin{figure}[h]
    \includegraphics[width=8cm]{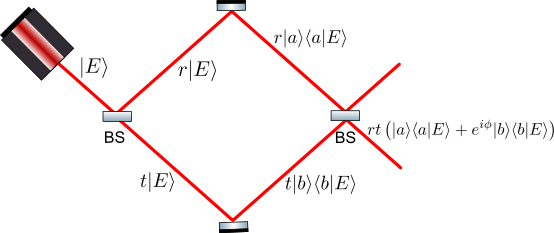}
    \caption{Scheme for a practical observation of Kirkwood-Dirac distribution for light polarization.\label{fig:first-scheme_measurement_polarization}}
\end{figure}

The scheme used to measure the Kirkwood-Dirac distribution for polarization is shown in Fig.~\ref{fig:first-scheme_measurement_polarization}. It consists of a Mach-Zehnder interferometer in which the beam splitters and mirrors act identically on all polarization components, their transmission and reflection coefficients, $t$ and $r$, are therefore polarization independent. In the upper arm of the interferometer, the field undergoes the transformation $|a\rangle\langle a|$, which can be implemented by a generalized polarizer projecting the beam onto the polarization state $|a\rangle$. Similarly, the lower arm implements the transformation $|b\rangle\langle b|$ by projecting onto the state $|b\rangle$. Such generalized polarizers can be realized experimentally by applying an appropriate unitary polarization transformation followed by a standard linear polarizer.

Considering all these elements, the complex amplitude of the field emerging from the lower output port of the beam splitter $|E^\prime \rangle$ can be expressed as\\
\begin{equation}
   |E^\prime \rangle = rt \left (|a\rangle\langle a|E\rangle + e^{i\phi}|b\rangle\langle b|E\rangle\right) ,
\end{equation}
where $\phi$ denotes the interferometric phase difference accumulated between the upper and lower arms of the interferometer. The corresponding field intensity is then given by 
\begin{eqnarray}
   & I=\langle E^\prime |E^\prime \rangle = |rt |^2 \left (  \langle E |a\rangle \langle a |E \rangle + \langle E |b \rangle \langle b |E \rangle \right . &\nonumber \\ & & \nonumber \\
   & +  \left . e^{i \phi}  \langle E |a \rangle \langle a | b \rangle \langle b |E \rangle + e^{-i \phi}  \langle E | b\rangle \langle b |a \rangle \langle a |E \rangle \right ) . &
\end{eqnarray}
Taking the ensemble average over the field realizations, the resulting expression can be expressed as
\begin{eqnarray}
  & I (\phi ) = |rt |^2 \left (  \langle a | \Gamma |a \rangle +\langle b | \Gamma |b \rangle   \right . & \nonumber \\ & & \nonumber \\
  & \left . + e^{i \phi} \langle a |b \rangle \langle b|\Gamma | a \rangle + e^{-i \phi}  \langle b |a \rangle \langle a |\Gamma |b \rangle \right ) &.
\end{eqnarray}
{}\\
By appropriately choosing the interferometric phase difference, one can, for example, obtain the Margenau-Hill distribution as the real part of the Kirkwood-Dirac distribution
\begin{equation}
   M(a,b)= \frac{1}{2} \left ( \langle a |\Gamma |b \rangle \langle b |a \rangle + \langle b|\Gamma | a \rangle \langle a |b \rangle \right ),
\end{equation}
which can be determined as
\begin{equation}
     M(a,b) = \frac{1}{4|rt|^2} \left [ I(0)-I(\pi ) \right ] .
\end{equation}
Moreover, the Kirkwood--Dirac distribution can be obtained by integrating over all possible values of the interferometric phase difference $\phi$: 
\begin{equation}
   \int_{0}^{2\pi} d\phi \, e^{i\phi} I(\phi) = 2 \pi  |rt|^2 K(a,b).
\end{equation} 
Experimentally, this quantity can be accessed by measuring $I(\phi)$ for several values of $\phi$ and performing the integration numerically. In practice, this is equivalent to summing over a discrete set of phase values.

\subsection{Interpretation of the anomalous values of the Kirkwood–Dirac distribution in polarization}

In its original introduction in the quantum domain the Kikwood-Dirac distribution is intended to be a joint distribution for pairs of observables. Thus, its complex or negative values are taken as signatures of nonclassical behavior, reflecting the fact that there is no way of assigning values to noncompatible observables.  

In the classical-optics domain, the Kirkwood-Dirac distribution, such as the better known Wigner function for example, no longer has a statistical meaning {\it per se}. Instead it indicates the way the field intensity is distributed along field variables, such as position and propagation direction regarding radiometry for example. So, complex or negative values for the classical Kirkwood-Dirac, as well as negative values for the Wigner function, may be referred to as {\it anomalous} or {\it strange} values, rather than {\it nonclassical}. In this subsection, we investigate and interpret the conditions under which the Kirkwood–Dirac  distribution displays anomalous values when applied to classical optical fields in the polarization degree of freedom.

One clear signature of anomalous behavior in the Kirkwood-Dirac distribution is the appearance of complex values, i.e., a nonvanishing imaginary part. This feature is often associated with the presence of coherences, namely the off-diagonal elements of the state $\Gamma$, these are the matrix elements $\Gamma_{1,2}$ and $\Gamma_{2,1}$ in Eq.~(\ref{MG}). Indeed, when coherences are present and the probe states $\ket{a}$ and $\ket{b}$ are distinct, the Kirkwood-Dirac quasiprobability distribution generally acquires a nonzero imaginary contribution. In this regard, in Ref.~\cite{BD23} it is shown the way the imaginary part of the distribution allows to obtain the coherences of a quantum state. 

However, the imaginary part of the Kirkwood-Dirac distribution does not exclusively quantify the magnitude of the coherences. Even in the absence of off-diagonal elements, $\Gamma_{1,2}=\Gamma_{2,1}=0$, there is a nonvanishing imaginary part. Let us show this in more detail. To this end we recall the general expression for $K(a,b)$ in Eq.~(\ref{Kab}) to get its imaginary part as 
\begin{equation}
    {\rm Im}\left [K(a,b) \right ] = \frac{ab}{4} \left ( \bm{s}_A \times \bm{s}_B \right )\cdot \bm{s} ,
\end{equation}
where we recall that $\bm{s}$ are the Stokes parameters of our field state in Eq. (\ref{GSp}). If there are no coherences, we have $s_x=s_y=0$, and then 
\begin{equation}
\label{eq:imaginary_part_of_the_distribution}
    {\rm Im}\left [K(a,b) \right ] = \frac{ab}{4} \left ( \bm{s}_A \times \bm{s}_B \right )_z \left ( \Gamma_{1,1} - \Gamma_{2,2} \right ),
\end{equation}
being 
\begin{equation}
\left ( \bm{s}_A \times \bm{s}_B \right )_z = -\sin \theta_A \sin \theta_B \cos \left (\varphi_A-\varphi_B \right ) .
\end{equation}

As Eq.~\eqref{eq:imaginary_part_of_the_distribution} shows, even when $\Gamma_{1,2}=\Gamma_{2,1}=0$, the distribution may still exhibit a nonzero imaginary contribution. This term depends both on the population imbalance $\Gamma_{1,1}-\Gamma_{2,2}$ and on the relative geometric configuration of the probe states on the Poincar\'e sphere. In particular, the imaginary part vanishes when the azimuthal angles coincide, $\varphi_A=\varphi_B$, or when the populations are balanced, $\Gamma_{1,1}=\Gamma_{2,2}=1/2$, this is $\Gamma \propto \mathcal{I} $. 

Therefore, in polarization systems, the imaginary component of the Kirkwood-Dirac quasiprobability distribution encodes both intrinsic properties of the state $\Gamma$ (coherences and population imbalance) and geometric properties of the chosen probe states. It thus reflects not only the characteristics of the state under investigation, but also the relative configuration of the measurement basis.

In this regard we may distinguish between {\it coherences}, these are the nondiagonal matrix elements in a given polarization basis, and the {\it total coherence}, that will refer to a basis independent measure of coherence. Actually, this role is played in the context of polarization by the degree of polarization, that can be conveniently described by the Hilbert-Schmidt distance of the normalized $\Gamma$ to the normalized identity $\mathcal{I} /2$ representing totally unpolarized light
\begin{equation}
    \mathcal{P}^2 = 2 \mathrm{tr} \left [ \left (\Gamma - \mathcal{I} /2 \right )^2 \right ] = 2 {\rm tr} \left ( \Gamma^2 \right ) - 1,
\end{equation}
where it is understood that intensity units are used so that $\mathrm{tr}\left( \Gamma\right) = 1$. Even in this case of degree of polarization as total coherence we may find a readily connection to the Kirkwood-Dirac distribution. To this end, let us start with the most general distribution in Eq.~(\ref{Kab}) to get 
\begin{eqnarray}
& \sum_{a,b} |K(a,b)|^2 = \frac{1}{4} \left \{ 1 + \left ( \bm{s}_A \cdot \bm{s}_B \right )^2+ \left ( \bm{s}_A \cdot \bm{s} \right )^2 \right . & \nonumber \\ & & \nonumber \\
& \left .+ \left ( \bm{s}_B \cdot \bm{s} \right )^2  + \left [ \left ( \bm{s}_A \times \bm{s}_B \right ) \cdot \bm{s} \right ]^2 \right \} . &
\end{eqnarray}
Choosing the polarization bases as orthogonal $\bm{s}_A \cdot \bm{s}_B=0$ we get 
\begin{equation}
    \sum_{a,b} |K(a,b)|^2 =\frac{1}{4} \left ( 1 + \mathcal{P}^2 \right ) = {\frac{1}{2} \rm tr} \left ( \Gamma^2 \right ).
\end{equation}

Thus, the sum of the squared moduli of the Kirkwood-Dirac distribution provides a meaningful quantifier of a basis-independent total coherence, which in this case manifests itself as the degree of polarization. In turn, this is essentially the deviation of $\Gamma$ from idempotency as mentioned above, this is the state purity in quantum physics.

In passing, note that the complex values on $K(a,b)$ arise from the correlations between the polarizations represented by $\ket{a}$ and $\ket{b}$. This is particularly intriguing, since in the quantum case correlations between incompatible observables, more precisely, observables that are not jointly measurable, do not admit a straightforward interpretation. The coherence meaning developed in this work in the classical case may be useful as well to understand these features in the quantum domain.

A second signature of anomalous behavior in the Kirkwood-Dirac distribution is the possible negativity of its real part, which is again often associated with the presence of coherences in the state. Indeed, the off-diagonal elements of $\Gamma$ contribute to the emergence of negative values in the real part of the Kirkwood-Dirac distribution. However, as in the case of the imaginary component, coherences are not the sole source of this behavior. To see this we extract the real part of $K(a,b)$ in Eq.~(\ref{Kab}) to get 
\begin{equation}
    {\rm Re}\left [K(a,b) \right ] = \frac{1}{4} \left [ 1+ ab \bm{s}_A \cdot \bm{s}_B + \left ( a \bm{s}_A + b \bm{s}_B \right )\cdot \bm{s} \right ] .
\end{equation}
When the coherences vanish, $\Gamma_{1,2}=\Gamma_{2,1}=0$, that is $s_x=s_y=0$, this real part takes the form
\begin{eqnarray}
\label{eq:real_part_KD}
  &  {\rm Re}\left [K(a,b) \right ] = \frac{1}{4} \left [ 1+ ab \bm{s}_A \cdot \bm{s}_B \right . & \nonumber \\ & & \nonumber \\ & \left . + \left ( a \bm{s}_A + b \bm{s}_B \right )_z \left ( \Gamma_{1,1} - \Gamma_{2,2} \right ) \right ], &
\end{eqnarray} 
where 
\begin{eqnarray}
& \bm{s}_A \cdot \bm{s}_B = \cos \theta_A \cos \theta_B+\sin \theta_A \sin \theta_B \cos \left (\varphi_A-\varphi_B \right ) , & \nonumber \\
& & \\
& \left ( a \bm{s}_A + b \bm{s}_B \right )_z = a \cos \theta_A  + b \cos \theta_B . &\nonumber
\end{eqnarray}
As Eq.~\eqref{eq:real_part_KD} shows, even in the absence of coherences, the real part contains several trigonometric contributions that may assume either positive or negative values. Consequently, negativity is here again governed by the population imbalance encoded in $\Gamma_{1,1}-\Gamma_{2,2}$ and by the geometric configuration of the probe states $\ket{a}$ and $\ket{b}$ on the Poincar\'e sphere. In particular, when the populations are balanced, $\Gamma_{1,1}=\Gamma_{2,2}=1/2$, the real part remains nonnegative. Therefore, the anomalous features of the Kirkwood--Dirac distribution arise not only from the presence of coherences, but also from the interplay between population imbalance and the relative geometry of the chosen measurement basis.
\section{Connection between the Kirkwood-Dirac distribution and noisy joint measurements}\label{section_4}

The polarization Kirkwood-Dirac distribution can be as well retrieved from experimental procedures addressing a noisy joint observation of the polarization variables in question, providing they contain all the information that would be required to obtain their exact values. This holds both in the quantum and classical realms. In this sense the structure of the Kirkwood-Dirac distribution closely resembles that arising in joint noisy measurements of complementary observables, as discussed for example in Refs.~\cite{MAL20,GBAL20,AL25}. For definiteness let us consider that the polarization bases correspond to polarization states being eigenvectors of the Pauli matrices $\hat\sigma_{x,y}$. Their Stokes parameters are $S_X=(1,0,0)$ and $S_Y=(0,1,0)$, so that Eq.~(\ref{Kab}) reads
\begin{equation}
\label{pc}
K(x,y) = \frac{1}{4}\left ( 1+ x s_x+ y s_y + i xy s_z \right ) ,
\end{equation}
In the classical case the noisy observation leads to a legitimate joint intensity distribution of the form
\begin{equation}
I(x,y)  = \frac{1}{4}\left ( 1+ x \gamma_x s_x+ y \gamma_y s_y + xy \gamma_{xy} s_z \right ) ,
\end{equation}
where $\gamma_x$, $\gamma_y$, and $\gamma_{xy}$ are positive parameters that quantify the additional noise unavoidably introduced by the measurement process. These coefficients satisfy the constraint
\begin{equation}
\gamma_x^2+\gamma_y^2+ \gamma_{xy}^2 \leq 1 ,
\end{equation}
which guarantees the non-negativity of $I(x,y)$. Suitable simple experimental implementations of such joint measurements can be found in Refs.~\cite{GBAL20,AL25}.

Within this framework, the Kirkwood-Dirac distribution $K(x,y)$ and the measurable joint distribution $I(x,y)$ can be related through a linear transformation defined by a kernel $\tilde{\chi}(x,y;x',y')$,
\begin{equation}
K(x,y) = \sum_{x^\prime,y^\prime} \tilde{\chi} (x,y;x^\prime, y^\prime ) I(x^\prime, y^\prime ) ,
\end{equation}
where
\begin{equation}
\tilde{\chi} (x,y;x^\prime, y^\prime )= \frac{1}{4} \left ( 1 +\frac{x x^\prime}{\gamma_x}+ \frac{y y^\prime}{\gamma_y} + i \frac{x y x^\prime y^\prime}{\gamma_{xy}} \right ) .
\end{equation}
This relation shows that the Kirkwood-Dirac distribution can be reconstructed from experimentally accessible joint noisy measurements of complementary observables. While $I(x,y)$ is a genuine, non-negative intensity distribution, the inversion encoded in the kernel $\tilde{\chi}$ effectively removes the measurement-induced noise, thereby revealing the underlying complex structure. In this sense, the Kirkwood-Dirac distribution provides a refined description of polarization coherence beyond what is directly accessible through standard joint measurements.

\section{Observation of the space-angular distribution}\label{section_5}

Next, we consider the classical-optics analog of the quantum coordinate-momentum Kirkwood-Dirac distribution $\langle x|\Gamma|p\rangle \langle p | x \rangle$. For definiteness, let us consider a one-dimensional problem with just a single transversal coordinate $x$. With regard to our original scheme in this case $a \rightarrow x$ so that $E(x) = \langle x | E \rangle$ represents the field strength at point $x$ while $b \rightarrow p$ so that $\tilde{E}(p) = \langle p | E \rangle$ represents the field strength in a plane-wave decomposition of the field, the one corresponding to the plane wave with wave-vector with transversal component  $k_x = k p$, where $k$ is the wavenumber. Then the classical version of $\langle x|\Gamma|p\rangle$ is 
\begin{equation}
    \Gamma(x,p) = \langle x | \Gamma |p \rangle = \left \langle E(x) \tilde{E}^\ast (p) \right \rangle. 
\end{equation}
In practical terms  $\tilde{E}(p)$ can be observed as a field strength at the back focal plane of a converging lens of focal length $f^\prime$, more specifically, in points $x^\prime$ of the back focal plane we have
\begin{equation}
    \tilde{E} (x^\prime) \propto \tilde{E} (p= x^\prime/f^\prime) .
\end{equation}
Moreover, we have $\langle p | x \rangle=\langle x | p \rangle^\ast$ with 
\begin{equation}
\label{xp}
    \langle x | p \rangle = \sqrt{\frac{k}{2\pi}} e^{ikpx} ,
\end{equation}
so that $k$ is the classical counterpart of the quantum $1/\hbar$.

In contrast to the polarization case, in principle we may address an interpretation of $K(x,p)$ in radiative terms, as it is often the case of, for example, the Wigner function. This is that $K(x,p)$ is intended to represent the light intensity leaving point $x$ in the direction specified by $p$. This will be addressed below in a simple Young interferometer. But before that, let us address the interpretation of $K(x,p)$ as generalized inter-modal coherence as we did in the polarization case. 

\subsection{Space-angular coherence}

\begin{figure}[h]
    \includegraphics[width=8cm]{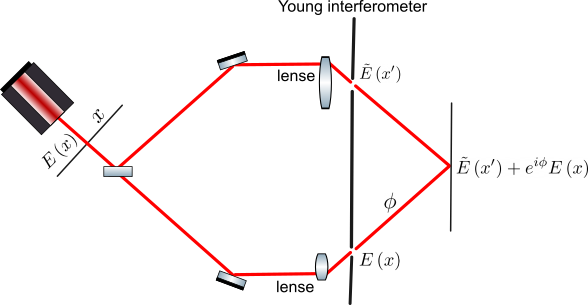}
    \caption{\label{fig:second_scheme}Scheme for a practical observation of Kirkwood-Dirac distribution for space-angular variables.}
\end{figure}

As in the polarization case, here we may find interesting to interpret $K(x,p)$ as a coherence function involving two different modal structures of the field, such as position and plane-waves. The proposed measuring scheme is illustrated in Fig.~\ref{fig:second_scheme}. The field state $|E\rangle$ has a given transverse profile $E(x)$ at the input plane. It is then split into two identical copies using a beam splitter. The upper and lower arms illuminate the apertures of a Young interferometer. In the upper arm, a converging lens is placed so that the upper aperture of the interferometer corresponds to a point $x^\prime$ in its back focal plane. Consequently, the complex field amplitude at the upper slit is proportional to $\tilde{E}(x^\prime)$. In the lower arm, a suitable lens forms an image of the input field on the plane of the Young interferometer, so that the complex amplitude at the lower slit is proportional to $E(x)$.

{}
Then, the field distribution at the interference plane is of the form 
\begin{equation}
   \tilde{E} (x^\prime) + e^{i \phi}  E(x)  ,
\end{equation}
where $\phi$ is the phase difference acquired by the two beams from the apertures to the observation point. The intensity distribution is then 
\begin{equation}
    |E(x)|^2 + | \tilde{E} (x^\prime)|^2 + e^{i \phi} \tilde{E}^\ast (x^\prime) E(x) +  e^{-i \phi} \tilde{E} (x^\prime) E^\ast (x) .
\end{equation}
After performing the ensemble average in Eq.~(\ref{enav}), we obtain
\begin{equation}
   I (\phi ) \propto \langle x |\Gamma | x \rangle + \langle p |\Gamma | p \rangle +  
   e^{i \phi}  \langle x |\Gamma | p \rangle + e^{-i \phi}  \langle p |\Gamma |x \rangle .
\end{equation}
From this point, we can proceed analogously to the polarization scheme to extract the Margenau-Hill and Kirkwood-Dirac distributions from $I(\phi)$ with suitable choices for $\phi$.

\subsection{Radiance and coherence}

Let us turn our attention to a picture where $\Gamma (x,p)$ exhibits its potential radiance nature as the intensity of light at a point $x$ propagating in the direction specified by $p$. This is classical complementarity in action, since wave behavior forbids a {\it bona fide} $\Gamma (x,p)$. To this end, the Young interferometer is a germane case study.

Let us consider two infinitesimal apertures located along the $x$ axis at the positions $x = \pm x_0$. The relevant values of the space-space mutual coherence function are then given by
\begin{eqnarray}
\label{Gaa}
& \Gamma (\pm x_0, \pm x_0 ) = \langle \pm x_0 |\Gamma |\pm x_0 \rangle = I_\pm, & \nonumber \\
 & & \\
 & \Gamma (x_0, - x_0 ) = \langle x_0 |\Gamma |- x_0 \rangle=\mu \sqrt{I_+ I_-} , & \nonumber
\end{eqnarray}
where $I_\pm$ denote the field intensities at each aperture, and $\mu$ is the complex degree of coherence, that we will assume to be real for the sake of simplicity. With these definitions, the space-angular Kirkwood-Dirac distribution $K(x,p)$ can be evaluated. We can start with the following identity that links $ K(\pm x_0,p)$ with the space-space mutual coherence function in Eq.~(\ref{Gaa}) 
\begin{eqnarray}
  &  K(\pm x_0,p) = \langle \pm x_0 |\Gamma | p \rangle \langle p | \pm x_0 \rangle & \nonumber \\ & & \nonumber \\ &= \langle p | \pm x_0 \rangle  \int dx^\prime \langle \pm x_0 |\Gamma | x^\prime \rangle \langle x^\prime | p \rangle . &
\end{eqnarray}
With this, we can take into account that there are only two nonvanishing contributions to $\langle \pm x_0 |\Gamma | x^\prime \rangle $. These are the corresponding to the aperture points $x^\prime = \pm x_0$. Hence, the Kirkwood-Dirac distribution takes the form
\begin{eqnarray}
   & K(\pm x_0,p) = \langle p | \pm x_0 \rangle  \left (  \langle \pm x_0 |\Gamma | x_0\rangle \langle x_0 | p \rangle \right . & \nonumber \\ & & \nonumber \\
 &  \left . +  \langle \pm x_0 |\Gamma | -x_0\rangle \langle - x_0 | p \rangle \right ). &
\end{eqnarray}
Taking Eq.~\ref{Gaa} into account, together with Eq.~(\ref{xp}) the Kirkwood-Dirac distribution can be expressed in terms of the $I_{\pm}$ intensities in Eq.~\ref{Gaa} as
\begin{equation}
    K(\pm x_0,p) = \frac{k}{2\pi} \left ( I_\pm + \mu \sqrt{I_+ I_-} e^{\mp 2ikpa} \right ) .
\end{equation}
This simple yet illustrative example shows in a different way that the anomalous values of the Kirkwood-Dirac distribution are directly linked to field coherence. A related feature appears in the Wigner function~\cite{AL06,AL07}, where coherence manifests itself through the emergence of negative values of $W(x,p)$ at positions $x$ where the field intensity vanishes, namely, at the midpoint between the apertures, that behaves as a secondary source of fictitious rays carrying negative intensity. 

It is also worth emphasizing that the real part of the Kirkwood-Dirac distribution coincides with the space-angle joint distribution derived in Ref.~\cite{GBAL20}, when the observation of complementary observables is addressed in classical and quantum optics. 

\subsection{Gauss-Schell beams}

Next, we present a further example to elucidate the connection between coherence and the complex character of the Kirkwood-Dirac distribution for a simple but very illustrative case of partially coherent Gaussian fields. A suitable expression for the mutual coherence function can be written as
\begin{equation}
\label{GS}
\Gamma (x,x^\prime) = \frac{1}{\sqrt{2 \pi \sigma^2}} \exp \left ( - \frac{x^2+x^{\prime 2}}{4 \sigma^2} \right ) \exp \left [ - \frac{(x- x^\prime)^2}{4 \mu^2} \right ] ,
\end{equation}
where for simplicity we neglect wavefront curvature assuming $\mu$ real and positive. The parameter $\sigma$ characterizes the beam width, and $\mu$ determines the transverse coherence length~\cite{MW95}. The intensity has been normalized such that $\int dx \, \Gamma(x,x) = 1$. 
The Kirkwood-Dirac distribution can then be computed as
\begin{equation}
    K(x,p) = \langle x |\Gamma | p \rangle \langle p | x \rangle = \langle p | x \rangle \int dx^\prime \langle  x^\prime |p \rangle \Gamma (x,x^\prime),
\end{equation}
leading to
\begin{equation}
\label{eq:KD_Gauss-Schell}
K(x,p)  = K_0 e^{-i\gamma kpx}e^{-\gamma_x x^2}e^{- \gamma_p k^2 p^2},
\end{equation}
where the coefficients are given by
\begin{eqnarray}
&\gamma = \frac{\mu^2}{\mu^2 + \sigma^2}, \quad \gamma_x = \frac{\mu^2 + 2\sigma^2}{4\sigma^2(\mu^2 + \sigma^2)}, \nonumber  \\ & & \\
& \gamma_p = \frac{\mu^2 \sigma^2}{\mu^2 + \sigma^2}, \quad K_0 = \frac{k}{\sqrt{2}\,\pi\sqrt{\mu^2 + \sigma^2}} . \nonumber 
\end{eqnarray}
It is immediately apparent from Eq.(~\ref{eq:KD_Gauss-Schell}) that the complex character of $K(x,p)$ arises from coherence effects. That is $K(x,p)$ is complex provided that $\mu \neq 0$.

\subsection{Interpretation of the anomalous values of the Kirkwood–Dirac distribution in the space–angular domain}\label{subsection:interpretation_continuous_variables}

Next, we analyze the nonclassical features of the Kirkwood-Dirac distribution in the continuous-variable space-angular domain case, $K(x,p) = \langle x|\Gamma|p\rangle\langle p|x\rangle$. This analysis includes the examination of the presence of negative regions in the real part, and the complex-valued nature of the distribution, all of which depend on both the state $\Gamma$ and the choice of measurement bases, here position and momentum.

The Kirkwood-Dirac distribution can generally be complex due to the intrinsic phase structure arising from the overlap between the chosen probe states, position and momentum. However, when the density operator is diagonal in the position basis,
\begin{equation}
\label{inc}
    \Gamma = \int dx\, \rho(x)\, |x\rangle\langle x|,
\end{equation}
the corresponding Kirkwood-Dirac distribution reduces to
\begin{align}
\label{eq:KD_diagonal_state}
K(x,p) &= \langle x|\Gamma|p\rangle \langle p|x\rangle
= \rho(x) \langle x|p\rangle \langle p|x\rangle \nonumber\\
&= \rho(x) \, |\langle x|p\rangle|^2
= \frac{k}{2\pi} \rho(x),
\end{align}
which is strictly real and nonnegative. We have adopted here the quantum view for being more intuitive. In such a case this incoherent state in Eq.~(\ref{inc}) must be understood as a suitable limit in which $|x\rangle$ are replaced by some physical states with a coordinate wave function centered at point $x$ with a vanishing width. On the other hand, from a classical perspective this is a field with no spatial coherence whatsoever.

In this case, the Kirkwood-Dirac distribution exhibits neither negative values nor complex components, despite the fact that the overlap $\langle x|p\rangle$ contains a phase factor. Classical optical fields that are diagonal in the position basis include thermal light and other forms of incoherent fields. Accordingly, only coherent light can display non-classical features in the Kirkwood-Dirac distribution. 

This behavior contrasts with the discrete-variable case, where even diagonal states can exhibit non-classical properties. This may be well regarded as another special feature of coherence regarding continuous variables, as the ones already found also in the quantum case for coherence in quadrature bases~\cite{AL23a}.

\begin{figure}[t]  
    \centering
\includegraphics[width= 9 cm]{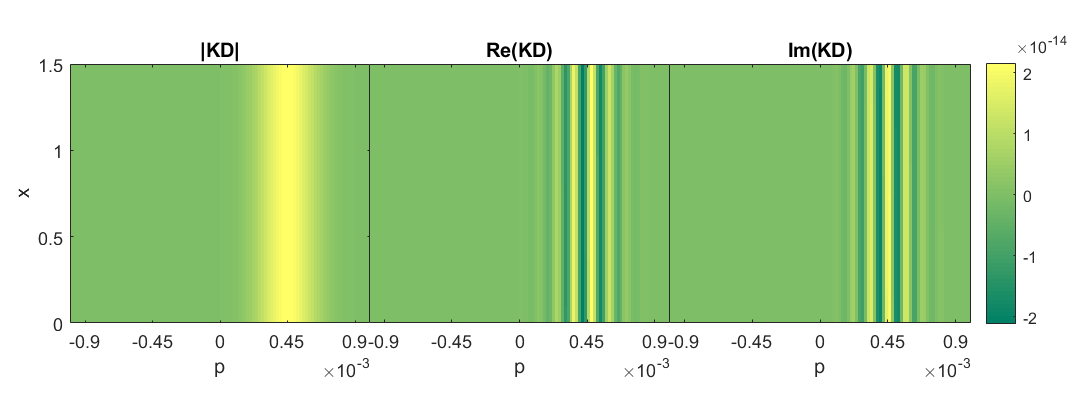} 
    \caption{Detail of the Kirkwood-Dirac distribution of a Gaussian optical beam centered at point $x_0 = 1$ with $\sigma = 0.03$. Shown are the modulus, real part, and imaginary part of the distribution in a relevant region of the plain $x-p$ revealing anomalous values. Although the beam is a simple classical Gaussian, the real part exhibits regions of negativity and the imaginary part is nonzero, highlighting that the Kirkwood-Dirac distribution can possess a rich complex structure even for classical coherent states.}
    \label{fig:KD_Gaussian}
\end{figure}

For the sake of illustration in Fig.~\ref{fig:KD_Gaussian} we consider the case of a fully coherent Gaussian field which, as expected, exhibits both negative values in the real part and a nonzero imaginary part due to its coherent nature.

We showed above in the polarization case that the sum of the square moduli of the Kirkwood-Dirac distribution provides a meaningful quantifier of a basis-independent total coherence, proportional to the state purity. We can readily show that the same holds in the space-angle case. This can be easily seen as
\begin{equation}
    \int dx dp |K(x,p) |^2 = \frac{k}{2\pi} \int dx dp \langle x | \Gamma | p \rangle \langle p | \Gamma | x \rangle ,
\end{equation}
that is
\begin{equation}
    \int dx dp |K(x,p) |^2 = \frac{k}{2\pi} \int dx dx^\prime \Gamma (x,x^\prime )\Gamma (x^\prime,x). 
\end{equation}
This is again 
\begin{equation}
    \int dx dp |K(x,p) |^2 = \frac{k}{2\pi} {\rm tr} \left ( \Gamma^2 \right ) ,
\end{equation}
as the classical analog of the equivalent result in the quantum domain
\begin{equation}
    \int dx dp |K(x,p) |^2 = \frac{1}{2\pi \hbar} {\rm tr} ( \rho^2 ), 
\end{equation}
is the state purity. This agrees well with equivalent results in the context of global coherence in classical optics~\cite{TSF04,AL07b}.

\section{Relation between Wigner and Kirkwood-Dirac distributions}\label{section_6}

In this section, we establish the connection between Kirkwood-Dirac distributions and Wigner distributions in two complementary settings: the space–angular domain, corresponding to a continuous-variable system, and the polarization domain, corresponding to a discrete-variable system.

\subsection{space-angular domain}

A suitable relation between the Wigner and Kirkwood-Dirac distributions can be derived from the following property of the Wigner function:
\begin{equation}
\langle x |\Gamma |p \rangle = \frac{2\pi}{k} \int dx^\prime dp^\prime W_{|p\rangle \langle x |} (x^\prime, p^\prime ) W_{\Gamma} (x^\prime, p^\prime )  
\end{equation}
where 
\begin{equation}
W_{\Gamma} \left(x, p\right) = \frac{k}{2\pi} \int dx^\prime \bra{ x-x^\prime/2} \Gamma \ket{ x+x^\prime/2 } e^{ikpx^\prime} .
\end{equation}
Likewise
\begin{equation}
W_{|p\rangle \langle x |} \left(x^\prime, p^\prime \right) = \frac{k}{2\pi} \int dy \langle{ x^\prime-y/2} \ket{p} \bra{x}  x^\prime + y/2 \rangle e^{ikpy} .
\end{equation}
Considering that $\langle x |x^\prime \rangle = \delta (x-x^\prime) $ and using Eq.~(\ref{xp}), we immediately obtain the desired relation in the form
\begin{eqnarray}
\label{Kxp}
& K(x,p) = \langle x |\Gamma |p \rangle \langle p |x\rangle  & \nonumber \\
 & & \nonumber \\
 &= \frac{k}{\pi} \int dx^\prime dp^\prime e^{-2ik(p-p^\prime)(x-x^\prime)} W_{\Gamma} (x^\prime, p^\prime ) , &
\end{eqnarray}
which fits with the inverse relation in Refs.~\cite{BSL10,MR14}.\\

Physically, this relation shows that $K(x,p)$ is a phase-space correlation between the position $x$ and momentum $p$, obtained as a Fourier-like transform of the Wigner function. The oscillatory kernel $e^{-2 i k (p - p^\prime)(x - x^\prime)}$ encodes interference between the classical phase-space coordinates $(x,p)$ and the points $(x^\prime,p^\prime)$ of the Wigner distribution. The complex nature of $K(x,p)$ arises from the oscillatory kernel $e^{-2 i k (p - p^\prime)(x - x^\prime)}$, which encodes interference between different phase-space points. 
\subsection{Polarization domain}

For a similar relation in the polarization case, let us again refer to two orthogonal polarization bases $|x\rangle$, $|y\rangle$ eigenstates of the Pauli matrices $\hat\sigma_x$ and $\hat\sigma_y$
with $x, y = \pm 1$. The states $|x = \pm 1\rangle$ may, for instance, represent horizontal and vertical linear polarization states. We may consider the discrete Wigner function introduced in Ref.~\cite{LP98} that can be expressed in this case as 
\begin{equation}
W(x,y)  = \frac{1}{4}\left ( 1+ x s_x+ y s_y + xy s_z \right ) .
\end{equation}
It can be seen that the corresponding Kirkwood-Dirac distribution $K(x,y)$ in Eq.~(\ref{pc}) and the Wigner distribution $W(x,y)$ are closely related, and can be connected through a linear transformation defined by a kernel $\chi(x,y;x',y')$:
\begin{equation}
K(x,y) = \sum_{x^\prime,y^\prime} \chi (x,y;x^\prime, y^\prime ) W(x^\prime, y^\prime ).
\end{equation}
The transformation kernel is given by
\begin{equation}
\chi (x,y;x^\prime, y^\prime )= \frac{1}{4} \left ( 1 +x x^\prime+y y^\prime + i x y x^\prime y^\prime \right ) .
\end{equation}
This relation shows that the Kirkwood-Dirac distribution can be understood as a complex representation obtained from the Wigner distribution through a nontrivial mixing of phase-space variables. While the Wigner distribution is real (though not necessarily positive), the kernel $\chi$ introduces an explicitly imaginary contribution proportional to $x y x' y'$.  In this sense, the complex character of $K(x,y)$ encodes additional phase information associated with noncommutativity, providing a representation that is particularly sensitive to interference effects.

\section{Conclusions}\label{section_8}

In this work, we have developed a thorough analysis of Kirkwood-Dirac distributions in the realm of classical optics. This analysis reveals a deep connection of this theoretical tool with a key concept in optics, namely coherence. This connection is already present in their definition, that can be regarded as a generalization of mutual coherence functions or polarization matrices by involving two different bases instead of one. 

Moreover, this coherence-based interpretation explains the so-called anomalous values of Kirkwood-Dirac distributions, which are classical-optics counterparts of nonclassical behavior in the quantum domain. We have shown that the anomalous values are readily connected with diverse manifestations of coherence in classical optics. This includes local coherence, in the form of nondiagonal matrix elements of the mutual coherence function in a given basis, as well as global coherence, represented by the deviation of the mutual coherence function from idempotency, as very well illustrated by the degree of polarization. We have found that this holds in all the cases considered in this work, namely, polarization, interference and wave propagation. We have found that anomalous values reflect not only the characteristics of the state under investigation, but also the relative configuration of the measurement basis.

Beyond the abstract analysis, we have proposed diverse methods of experimental determination of these distributions. Essentially, they are all based on interference, in agreement with the coherence meaning of the distributions. Moreover, we have found suitable relations with noisy joint measurements as well as with the much better known and widely used Wigner function. The key point being that they share a common structure. 

These results may open new avenues for research in classical optics that will benefit from the research already done on this distribution in the quantum domain. This may be well extended beyond the examples addressed in this work by considering some different combinations of field variables, mixing polarization, interference and propagation properties for example. This might be the case of the much debated issue of coherence in the interference of partially polarized waves~\cite{BK63,EW03,TSF03,STF04,RG05,AL07a}. 

Conversely, the coherence-based interpretation of Kirkwood-Dirac distributions in the classical case may be rather useful in the quantum domain. This is because coherence is being revealed as the basic resource for quantum properties and their practical application. For example, the analysis in this work provides an interferometric-coherence meaning alternative to the more standard interpretation as a joint distribution of complementary variables. In this sense, the nonclassical behavior of quantum Kirkwood-Dirac distributions is naturally understood when they are regarded as the ability to interfere in the generalized framework defined by the two bases considered. 

\section*{Acknowledgments}
This research was supported by the EUTOPIA Science and Innovation Fellowship Programme
and funded by the European Union Horizon 2020 programme under theMarie Skłodowska-Curie Grant Agreement No. 945380.


\begin{thebibliography}{00}

\bibitem{Dirac45}
P.A.M. Dirac, On the analogy between classical and quantum mechanics, \href{https://journals.aps.org/rmp/abstract/10.1103/RevModPhys.17.195}{Rev. Mod. Phys. {\bf 17}, 195 (1945).}

\bibitem{Kirkwood33}
J.G. Kirkwood, Quantum statistics of almost classical assemblies, \href{https://journals.aps.org/pr/abstract/10.1103/PhysRev.44.31}{Phys. Rev. {\bf 44}, 31 (1933).}

\bibitem{Wigner32}
E. Wigner, On the quantum correction for thermodynamic equilibrium, 
\href{https://journals.aps.org/pr/abstract/10.1103/PhysRev.40.749}{Phys. Rev. {\bf 40}, 749 (1932).}

\bibitem{CahillGlauber69}
K.E. Cahill and R.J. Glauber, Ordered expansions in boson amplitude operators, 
\href{https://journals.aps.org/pr/abstract/10.1103/PhysRev.177.1857}{Phys. Rev. {\bf 177}, 1857 (1969).}

\bibitem{AAV88}
Y. Aharonov, D.Z. Albert, and L. Vaidman, How the result of a measurement of a component of the spin of a spin-1/2 particle can turn out to be 100, 
\href{https://journals.aps.org/prl/abstract/10.1103/PhysRevLett.60.1351}
{Phys. Rev. Lett. {\bf 60}, 1351 (1988).}

\bibitem{Wagner23}
R. Wagner, Z. Schwartzman-Nowik, I.L. Paiva, et al., 
Quantum circuits for measuring weak values, Kirkwood–Dirac quasiprobability distributions, and state spectra, 
\href{https://doi.org/10.1088/2058-9565/ad124c}{Quantum Sci. Technol. {\bf 9}, 015030 (2024).}

\bibitem{Lostaglio23}
M. Lostaglio, A. Belenchia, A. Levy, S. Hernández-Gómez, N. Fabbri, and S. Gherardini, 
Kirkwood-Dirac quasiprobability approach to the statistics of incompatible observables, 
\href{https://doi.org/10.22331/q-2023-10-09-1128}{Quantum {\bf 7}, 1128 (2023).}


\bibitem{ArvidssonShukur24}
D.R.M. Arvidsson-Shukur, W.F. Braasch Jr, S. De Bièvre, J. Dressel, A.N. Jordan, C. Langrenez, M. Lostaglio, J.S. Lundeen, and N.Y. Halpern, 
Properties and applications of the Kirkwood–Dirac distribution, 
\href{https://iopscience.iop.org/article/10.1088/1367-2630/ada05d}{New J. Phys. {\bf 26}, 121201 (2024).}

\bibitem{HeFu23}
J. He and S. Fu, Nonclassicality of the Kirkwood–Dirac quasiprobability distribution via quantum modification terms,
\href{https://doi.org/10.1103/PhysRevA.109.012215}{Phys. Rev. A {\bf 109}, 012215 (2024).}

\bibitem{Ferrie11}
C. Ferrie, Quasi-probability representations of quantum theory with applications to quantum information science, 
\href{https://iopscience.iop.org/article/10.1088/0034-4885/74/11/116001}{Rep. Prog. Phys. {\bf 74}, 116001 (2011).}


\bibitem{Hernandez-Gomez24}
S. Hernández-Gómez, T. Isogawa, A. Belenchia, A. Levy, N. Fabbri, and S. Gherardini,
Interferometry of quantum correlation functions to access quasiprobability distribution of work, \href{https://doi.org/10.1038/s41534-024-00913-x}{npj Quantum Information {\bf 10}, 112 (2024).}


\bibitem{LRBWT99} 
K. F. Lee, F. Reil, S. Bali, A. Wax, and J. E. Thomas, Heterodyne measurement of Wigner distributions
for classical optical fields, \href{https://doi.org/10.1364/OL.24.001370}{Opt. Lett. {\bf 24}, 1370 (1999)}.

\bibitem{BSL10}
V. Bollen, Y. M. Sua, and K. F. Lee, Direct measurement of the Kirkwood-Rihaczek distribution for the spatial properties of a coherent light beam, \href{https://doi.org/10.1103/PhysRevA.81.063826 }{Phys. Rev. A {\bf 81}, 063826 (2010)}.

\bibitem{AL22}
L. Ares and A. Luis, Distance-based approach to quantum coherence and nonclassicality, 
\href{https://doi.org/10.1103/PhysRevA.106.012415}{Phys. Rev. A {\bf 106}, 012415 (2022)}.

\bibitem{BD23}
A. Budiyono and H. K. Dipojono, Quantifying quantum coherence via Kirkwood-Dirac quasiprobability,
\href{https://doi.org/ 10.1103/PhysRevA.107.022408}{Phys. Rev. A {\bf 107}, 022408 (2023)}.

\bibitem{MAL20}
E. Masa, L. Ares, and A. Luis, Nonclassical joint distributions and Bell measurements, \href{https://doi.org/10.1016/j.physleta.2020.126416}{Phys. Lett. A {\bf 384}, 126416 (2020)}.

\bibitem{GBAL20}
R. Galazo, I. Bartolom\'e, L. Ares, and A. Luis, Classical and quantum complementarity, \href{https://doi.org/10.1016/j.physleta.2020.126849}{Phys. Lett. A {\bf 384}, 126849 (2020)}.

\bibitem{AL25}
A. Luis, Generalized measurements for Bell tests in different probability spaces, \href{https://doi.org/10.48550/arXiv.2506.07496}{arXiv:2506.07496 [quant-ph]}.

\bibitem{AL06} 
A. Luis, Negativity, diﬀraction and interference for nongeometrical waves,
\href{https://doi.org/10.1016/j.optcom.2006.05.036}{Opt. Commun. {\bf 266}, 426 (2006).}



\bibitem{AL07} 
A. Luis, Complementary Huygens principle for geometrical and nongeometrical optics, \href{ https://doi.org/10.1088/0143-0807/28/2/008}{Eur. J. Phys. {\bf 28}, 231 (2007).}

\bibitem{MW95} 
L. Mandel and E. Wolf, {\it Optical Coherence and Quantum Optics} (Cambridge University Press, Cambridge, U.K., 1995) 

\bibitem{AL23a} 
L. Ares and A. Luis, Coherence and incoherence in quadrature basis,  \href{https://doi.org/10.48550/arXiv.2307.08333}{arXiv:2307.08333 [quant-ph]}

\bibitem{TSF04}
J. Tervo, T. Set\"al\"a, and A. T. Friberg, Theory of partially coherent electromagnetic fields in the space--frequency domain, \href{https://doi.org/10.1364/JOSAA.21.002205}{J. Opt. Soc. Am. A {\bf 21}, 2205 (2004)}. 

\bibitem{AL07b}
A. Luis, Overall degree of coherence for vectorial electromagnetic fields and the Wigner function,
\href{https://doi.org/10.1364/JOSAA.24.002070}{J. Opt. Soc. Am. A {\bf 24}, 2070 (2007)}.

\bibitem{MR14}
P. A. Mello and M. Revzen, Wigner function and the successive measurement of position and momentum
\href{https://doi.org/10.1103/PhysRevA.89.012106}{Phys. Rev. A {\bf 89}, 012106 (2014)}.

\bibitem{LP98}
A. Luis and J. Pe\v{r}ina, Discrete Wigner function for finite-dimensional systems,
\href{https://doi.org/10.1088/0305-4470/31/5/012}{J. Phys. A {\bf 31}, 1423 (1998)}.

\bibitem{BK63}
B. Karczewski, Degree of coherence of the electromagnetic field,
\href{https://doi.org/10.1016/S0375-9601(63)96329-1}{Phys. Lett.{\bf 5}, 191--192 (1963)}.

\bibitem{EW03}
E. Wolf, Unified theory of coherence and polarization of random electromagnetic beams 
\href{https://doi.org/10.1016/S0375-9601(03)00684-4}{Phys. Lett. A {\bf 312}, 263--267, (2003)}.

\bibitem{TSF03}
J. Tervo, T. Set\"al\"a, and A. T. Friberg, Degree of coherence for electromagnetic fields, \href{https://doi.org/10.1364/OE.11.001137}{Opt. Express {\bf 11}, 1137--1143 (2003)}.

\bibitem{STF04}
T. Set\"al\"a, J. Tervo, and A. T. Friberg, Complete electromagnetic coherence in the space--frequency domain, \href{https://doi.org/10.1364/OL.29.000328}{Opt. Lett. {\bf 29}, 328--330 (2004)}.

\bibitem{RG05}
P. R\'efr\'egier and F. Goudail, Invariant degrees of coherence of partially polarized light,  \href{https://doi.org/10.1364/OPEX.13.006051}{Opt. Express {\bf 13}, 6051 (2005).}

\bibitem{AL07a}
A. Luis, Degree of coherence for vectorial electromagnetic fields as the distance between correlation matrices, \href{https://doi.org/10.1364/JOSAA.24.001063}{J. Opt. Soc. Am. A {\bf 24}, 1063 (2007)}.

\end{thebibliography}
\end{document}